\documentclass[twocolumn,10pt]{tsfp9}
\usepackage{flushend}
\usepackage{graphicx}
\usepackage[authoryear,round]{natbib}

\title{Bypass transition and subcritical turbulence in plane Poiseuille flow}

\author{Stefan Zammert and Bruno Eckhardt
    \affiliation{
	Fachbereich Physik, 
	Philipps-Universit\"{a}t Marburg\\
	Renthof 6, D-35032 Marburg, Germany\\
    Stefan.Zammert@physik.uni-marburg.de,
    Bruno.Eckhardt@physik.uni-marburg.de
    }	
}

\begin{document}

\maketitle   

\fontsize{9}{11}\selectfont

\section*{ABSTRACT}
Plane Poiseuille flow shows turbulence at a Reynolds number that is lower than the
critical one for the onset of Tollmien-Schlichting waves. The transition to turbulence follows
the same route as the by-pass transition in boundary layers, i.e. finite amplitude perturbations
are required and the flow is dominated by downstream vortices and streaks in the transitional
regime. In order to relate the phenomenology in plane Poiseuille flow to our previous
studies of plane Couette flow \citep{Kreilos2012}, we study 
a symmetric subspace of plane Poiseuille flow in which the bifurcation cascade stands out 
clearly. By tracing the edge state, which in this system is a travelling wave, and its bifurcations, 
we can trace the formation of  a  chaotic attractor, the interior crisis that increase the
phase space volume affected by the flow, and the ultimate transition into a chaotic saddle
in a crisis bifurcation. After the boundary crisis we can observe transient chaos with 
exponentially distributed lifetimes. 

\section*{Introduction}
The laminar state of plane Poiseuille flow (PPF) becomes linearly unstable at a Reynolds number of $5772.22$ \citep{Orszag1971}.
Above this critical Reynolds number any small disturbance lead to the formation of
Tollmien-Schlichting (TS) waves. As time progresses, these waves grow in amplitude until they 
finally undergo a secondary instability, break up, and a reach a turbulent state.
Below the critical Reynolds number, just as for pipe and plane Couette flow,
turbulence can be observed down to Reynolds numbers between 
$600$ and $1000$ \citep{Carlson1982,Tuckerman2014,Xiong2015} 
although the laminar state is linearly stable. 
Moreover, while the transition mechanism via TS-waves is subcritical, the subcritical branch does not reach 
sufficiently low to explain the transition in that range. Therefore, the transition mechanism has to
be different from that associated with TS-waves.

\begin{figure}[b]
\centering
\includegraphics[]{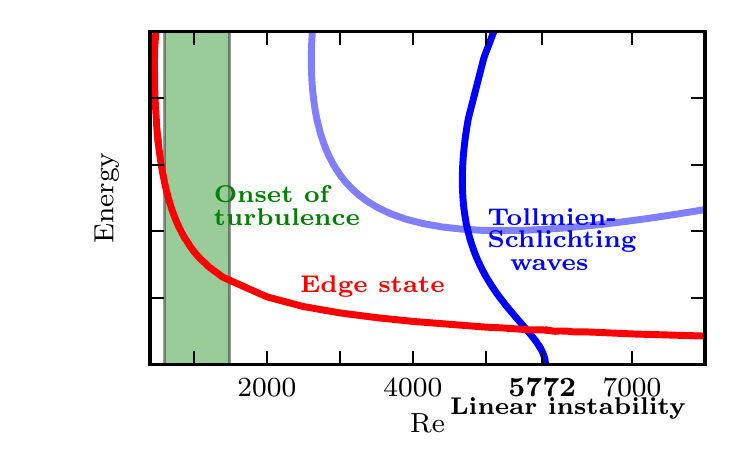}
\caption{Bifurcation diagram of plane Poiseuille flow. The laminar state becomes unstable to TS-waves in a
subcritical bifurcation at $Re=5772.22$.  The TS-waves exist down to $Re\approx 2600$ for the optimum wavelength (light blue). At low  Reynolds numbers the edge state appears and bypass-transition 
is possible for finite amplitude perturbations. The critical Reynolds number for the appearance of exact solutions is typically lower than the Reynolds number where turbulence is observed in experiments,
because near the critical point only a small part of state space is affected.
 \label{fig_PPFsystem}}
\end{figure}

Much insight into the process can be gained by studying the critical amplitudes that have to be
surpassed in order to trigger transition. They depend on Reynolds number and decrease with increasing Reynolds number \citep{Lemoult2012}. More importantly, the boundary between initial conditions that decay and those that 
reach turbulence is formed by states that are on the stable manifold of an intermediate state, the
edge state of the system \citep{Skufca2006}. The different states and their variation
with Reynolds number are shown in figure \ref{fig_PPFsystem}. 

The edge state also provides access to other coherent structures relevant for the dynamics. Tracing
it to lower Reynolds number one reaches the point where the state appears in a saddle-node
like bifurcation. Following the upper branch back to higher Reynolds numbers then gives states that
characterize the chaotic dynamics, i.e. the transition from simple exact coherent structures to more
complicated ones, very often in standard bifurcation such as Hopf-bifurcations and period doubling
cascades. This connection between exact coherent structures and the temporal properties of the
chaotic dynamics was studied in plane Couette \citep{Kreilos2012} and pipe flow \citep{Avila2013}. 
For the case of plane Couette flow it was found that the upper branch of the Nagata-Busse-Clever-solution,
the first three-dimensional equilibrium solution identified for plane Couette flow \citep{Nagata1990,Clever1992}, 
undergoes a series of bifurcations leading to a chaotic attractor. A boundary crisis bifurcation 
\citep{Lai2011} turns the attractor into a chaotic saddle. Following the boundary crisis, the lifetimes are 
exponentially distributes, as observed in experiments  \citep{Bottin1998} and simulations \citep{Schneider2010b}.
In pipe flow the same mechanism exist. Here, the relevant state that is the starting point of the cascade
is a streamwise localized periodic orbit which can be found using the technique of edge tracking \citep{Skufca2006}
in a symmetric subspace \citep{Avila2013}.

We recently found the same mechanism for the generation of subcritical 
turbulence in PPF  \citep{Zammert2015}. For a symmetric subspace of PPF we will 
present the edge state of the system and discuss the formation of the chaotic saddle. Furthermore,
we will show that attractors and saddles with quite different spatial wavelength exist in the state space
and discuss possible implications of this observation.

For our numerical simulation of plane Poiseuille flow
we use the Open-Source $\textit{Channelflow}$-code developed by J.F.Gibson \citep{J.F.Gibson2012}.
The code uses periodic boundary conditions in streamwise and spanwise direction together with
no-slip boundary conditions at both walls. A constant mass-flux is enforced in all our calculation.
The definition of the Reynolds number  $Re=U_{0}d \ \nu$ in this work uses half the distance between 
the plates $d$ and the center-line velocity $U_{0}$ of the laminar profile.

\section*{The edge state in a symmetric subspace}
We study PPF with two imposed symmetries
in a small periodic domain with a streamwise and spanwise wavelengths of $2\pi$ and $\pi$, respectively.
The used numerical resolution is $N_{x}\times N_{y} \times N_{z}=48 \times 65 \times 48$.
The imposed symmetries are a reflection at the center-plane,
\begin{equation}
s_{y}: [u,v,w](x,y,z)=[u,-v,w](x,-y,z),
\end{equation}
and a reflections in spanwise direction,
\begin{equation}
s_{z}: [u,v,w](x,y,z)=[u,v,-w](x,y,-z).
\end{equation}
In this symmetric subspace the dynamics is simpler
and exact coherent structures within this subspace are
stabilized, since all unstable direction pointing out of the subspace are removed.

We recently showed \citep{Zammert2015} that in this symmetry restricted subspace
the edge state, the attracting state on the boundary between laminar and turbulent initial conditions, is
a travelling wave, henceforth referred to as $TW_{Eyz}$. This travelling wave can be identified using edge tracking\citep{Skufca2006}.
The technique uses a bisection in amplitude to find trajectories on both sides of the laminar-turbulent boundary as approximation to
a trajectory on the laminar turbulent boundary. To stay close to the laminar-turbulent boundary from time to time
refinements by additional bisections are necessary. Using such refinement steps, edge tracking allows to follow trajectories
on the laminar turbulent boundary for an arbitrary long time and thus to find attracting states in the boundary.

\begin{figure}[]
\centering
\includegraphics[]{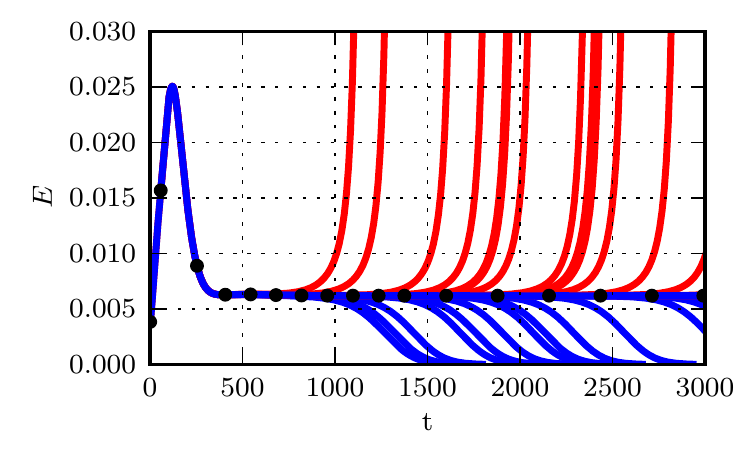}
\caption{Edge trajectory at $Re=2000$ in the symmetric subspace. For each refinement the red and blue lines are used for the trajectories on the turbulent and laminar side, respectively.
Black circles mark the positions of refinement steps, additional bisections, necessary to stay close to the trajectory on the edge.
 \label{fig_ET}}
\end{figure}

In figure \ref{fig_ET} the edge tracking used to find $TW_{Eyz}$ is shown.
The edge trajectory, approximated by trajectories on the laminar (blue) and the turbulent (red) side of the edge quickly reaches a state of constant energy 
\begin{equation}
 E(\vec{u})=\frac{1}{2 L_{x} L_{z}} \int_{0}^{L_{x}} \int_{-1}^{1}\int_{0}^{L_{z}}\vec{u}^{2}  dx \  dy \ dz .
\end{equation}
This indicates that the attracting state is a relative equilibrium. Using an initial condition on the edge trajectory as initial
guess, a Newton-search for a relative equilibrium converges to the travelling wave $TW_{Eyz}$

\begin{figure}[]
\centering
\includegraphics[]{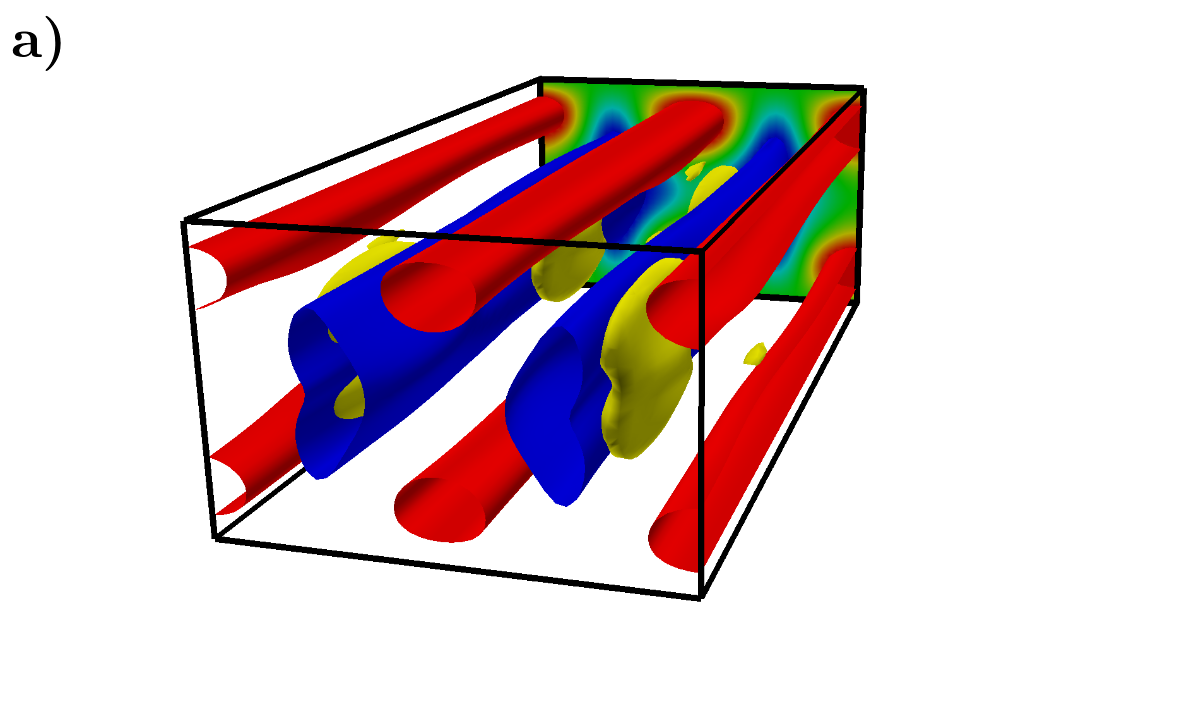}
\\
\includegraphics[]{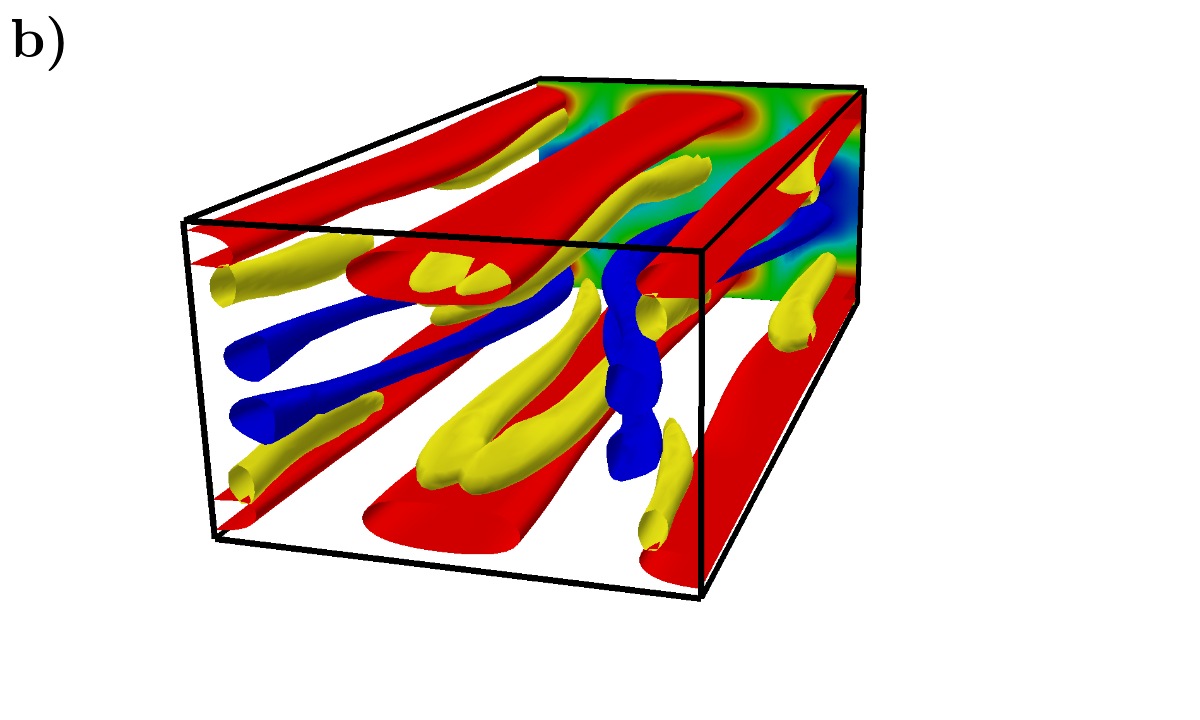}
\caption{Visualizations of the lower (a) and upper branch (b) of
the travelling wave $TW_{Eyz}$ at $Re=850$. The red and blue surfaces show iso-surfaces of $u=\pm0.15$ for the lower and 
 $u=\pm0.30$ for the upper branch. The yellow surfaces show iso-surfaces of the Q-vortex criterion.
 The iso-values are $0.008$ and $0.08$, respectively. On the back side of the box the
 streamwise velocity (deviation from laminar) is color coded. The vortices in the lower branch 
 are fairly straight with weak modulations in spanwise direction. The upper branch has
 a more complex structure.
 \label{fig_YZ_TWEyz}}
\end{figure}

The wave can be tracked in Reynolds number using a numerical continuation technique what allows to find its turning point and the upper branch can.
Visualizations of the lower and upper branch of the traveling wave are shown in figure \ref{fig_YZ_TWEyz}. 
The visualizations using iso-contours of the streamwise velocity (deviation form laminar) and the Q-vortex criterion \citep{Hussain95}
show that the state consists of streamwise streaks and vortices. On the lower branch the vortices are localed near the center-line
while on the upper branch the strongest vortices can be found closer to the walls. Furthermore, iso-contours of the Q-vortex criterion 
suggest that on the upper branch the  vortices have a hairpin-like structure because the tips of two
counter-rotating vortex tubes connect.

\section*{The formation of a chaotic saddle}
\begin{figure}[]
\centering
\includegraphics[width=0.4\textwidth]{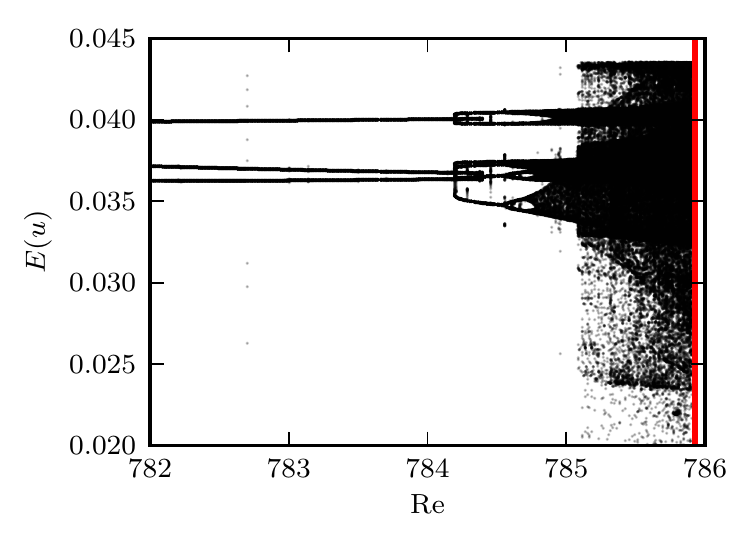}
\caption{Bifurcation diagram of the attractor in a computational domains with $L_{x}=2\pi$ and $L_{z}=1\pi$.
The states are visualized by plotting minima and maxima of $E(u)$ along trajectories. Continues lines then indicate
states with a simple periodicity, splittings (as near 784) an increase in period and complexity, and clouds
an irregular temporal variation. The red line markes the Reynolds number where the fluctuations in the state
meet the lower branch state and the attractor breaks up into a transient chaotic saddle.
 \label{fig_BifDiagLx2pi}}
\end{figure}

A series of bifurcations that start at the upper branch of the state  at $Re=707$ creates a chaotic attractor. 
The series starts with the bifurcation of a relative periodic orbit and involves
different kinds of bifurcations, e.g. period doublings and interior crisis bifurcations \citep{Grebogi1982,Lai2011}. 
Ultimately, the attractor is converted into a chaotic saddle by a boundary crisis bifurcation.
Sampling the minima and maxima of the energy
along trajectories on the attractor it is possible
to visualize the cascade in a bifurcation diagram. In figure \ref{fig_BifDiagLx2pi}
the diagram is shown for Reynolds numbers close to boundary crisis.

To investigate the influence of the boundary crisis on the state space structure,
two-dimensional slices of the state space are regarded.
Using two parameters, $\alpha$ and $A$, and the flow fields of the lower ($\vec{u}_{1}$) and upper ($\vec{u}_{2}$) branch of $TW_{Eyz}$
a slice of the state space can be defined by: 
\begin{equation}
{\vec{u}}(\alpha,A)=
A \frac{ (1-\alpha) {\vec{u}}_{1} + \alpha {\vec{u}}_{2}}{\sqrt{E\left((1-\alpha) {\vec{u}}_{1} + 
\alpha {\vec{u}}_{2}\right)}}.
\end{equation}
The parameter $\alpha$ allows for interpolation between the lower and upper branch and $A$ is the amplitude of the
flow field. 

\begin{figure}
\centering
\includegraphics[width=0.4\textwidth]{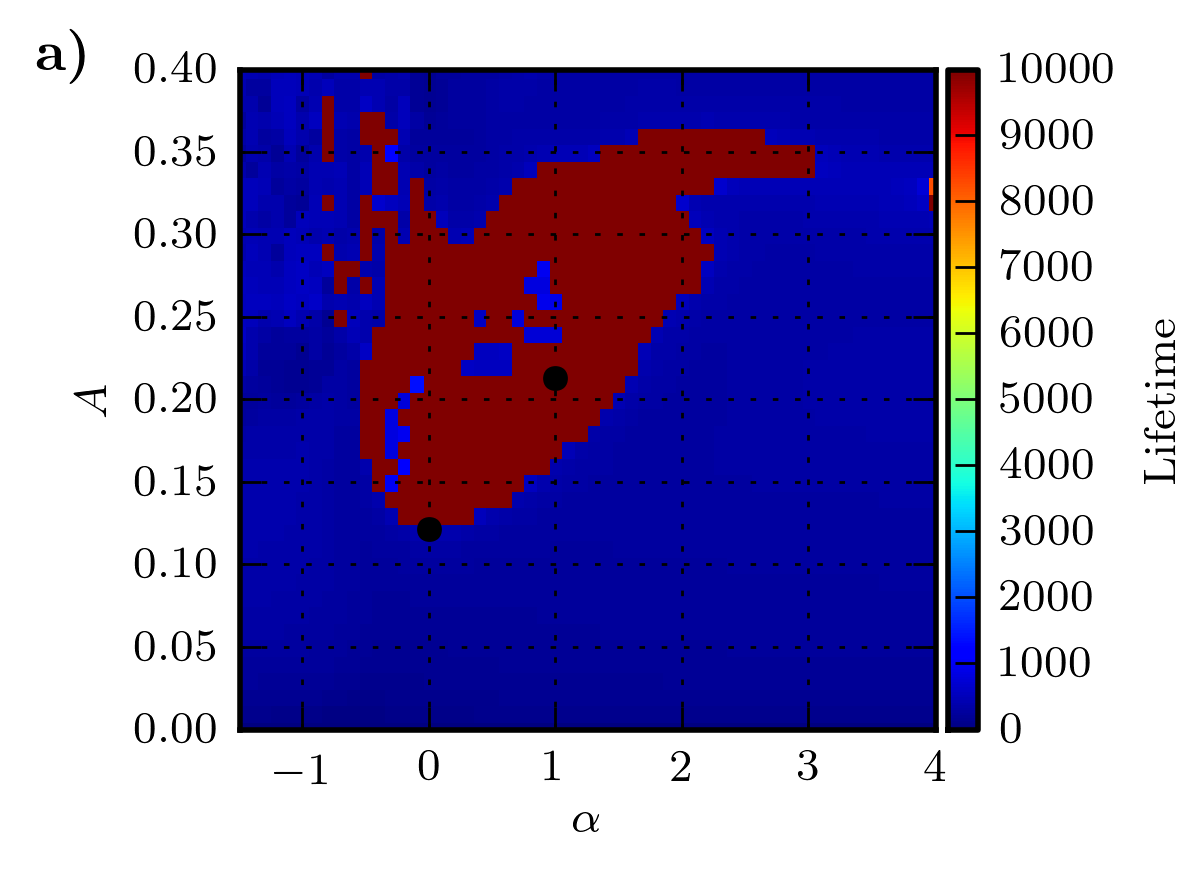}
\\
\includegraphics[width=0.4\textwidth]{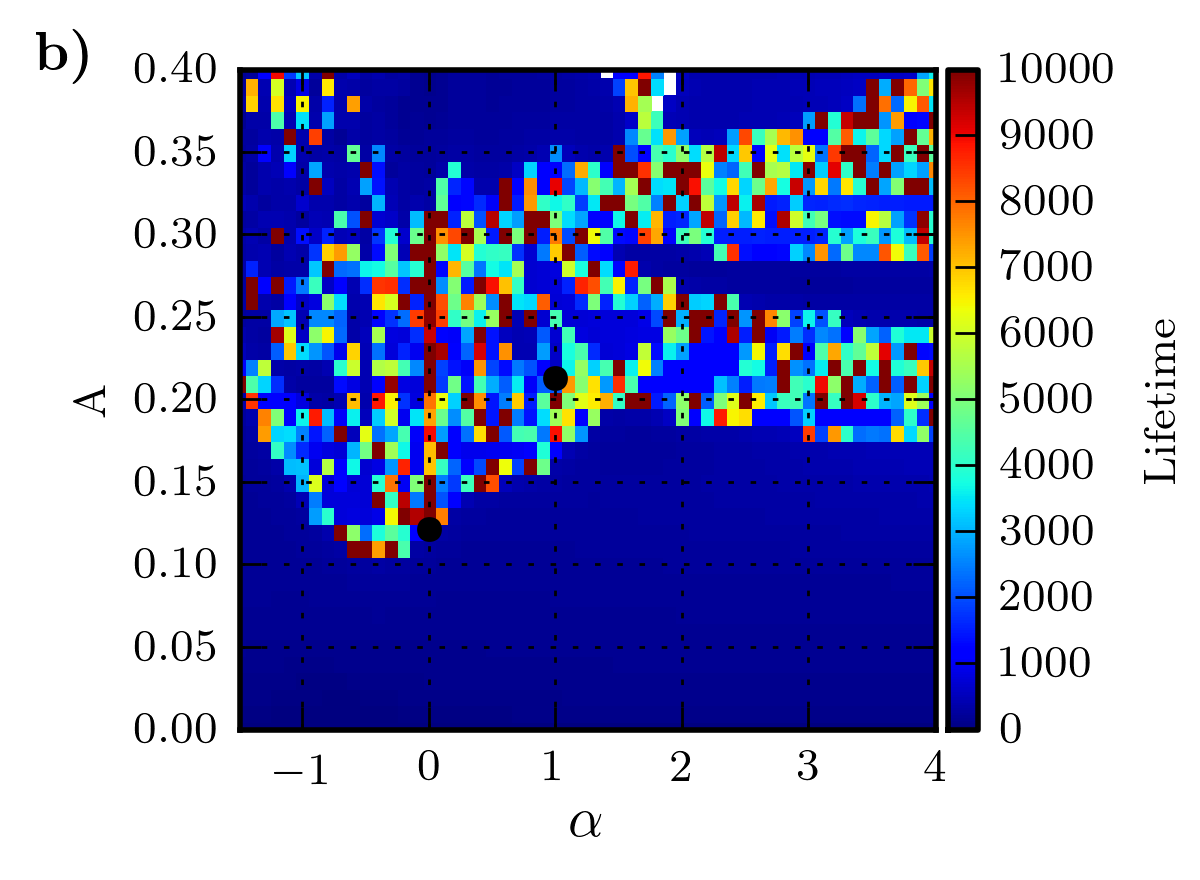}
\caption{State space slices for different Reynolds numbers. The initial conditions are color coded according to their lifetime, 
the time they need to fall below $10^{-4}$ in energy. Initial condition that do not fall below this
value within the observation time of $10^{4}$ time units are shown with this value as lifetime. 
For a) and b) the Reynolds numbers are 780 and 790, respectively. The positions of the upper and
lower branch of $TW_{Eyz}$ are marked with black dots.}
\label{fig_Basin}    
\end{figure}

Slices for two different Reynolds number are shown in figure \ref{fig_Basin}.
In these slices the initial conditions are color-coded according to the time needed to approach the laminar state.
(In practical we use the time the initial condition needs to fall below an certain very low value in energy.)
Below the boundary crisis (figure \ref{fig_Basin}a)  initial condition are either captured by the attractor (dark red) or return immediately to the laminar state (blue).
After the crisis (figure \ref{fig_Basin}b) the lifetime landscape is fractal as also observed in earlier investigations of plane Couette flow  by
 \cite{Schmiegel1997} and \cite{Kreilos2012}.The distribution of the lifetimes is exponential with a characteristic timescale that decreases with increasing Reynolds number 
and diverges with approaching the Reynolds number of the crisis.

It should be remarked that the wave $TW_{Eyz}$ does not connect to the Tollmien-Schlichting waves bifurcating from the laminar state.
Thus, the occurrence of subcritical turbulence in PPF is independent of the linear instability and completely analogous to the case of plane Couette or pipe flow.

\section*{Attractors and saddles with different wavelength}
The attractor and the saddle discussed in the previous section
have a streamwise  and spanwise  wavelength of $2\pi$ and $1\pi$, respectively.
The Reynolds numbers where the attractor appears and where it is converted into a chaotic saddle
depend, of course, on these spatial wavelengths. However, the phenomenon itself is quite robust to 
changes of these wavelengths and can be observed over a larger range of parameter values.

\begin{figure}[]
\centering
\includegraphics[width=0.4\textwidth]{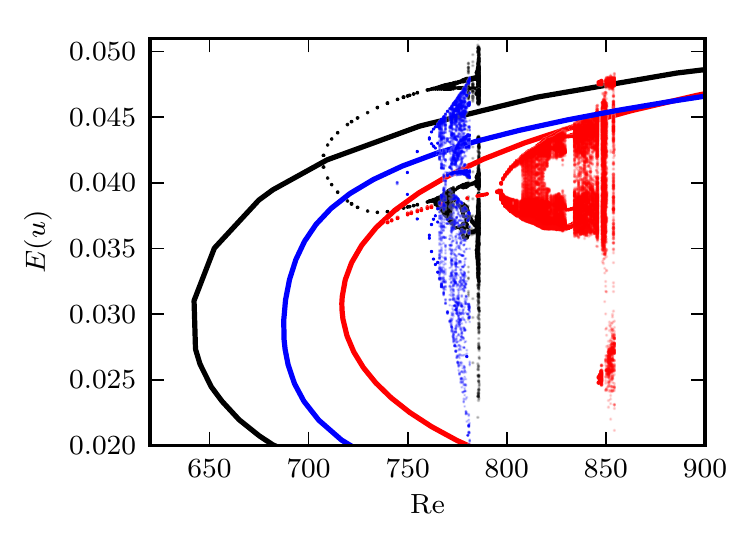}
\caption{Bifurcation diagrams for states with spatial wavelength of $3\pi$ (red), $2\pi$ (black) and $1.5\pi$ (blue) .
A solid line is used for the travelling wave $TW_{Eyz}$. 
The time dependent states are visualized by plotting minima and maxima of $E(u)$ along trajectories.
 \label{fig_AttraktorLx3pi}}
\end{figure}

For  streamwise wavelengths of $1.5\pi$ and $3\pi$ we study the attractors more detailed.
In figure \ref{fig_AttraktorLx3pi} both attractors are shown together with the one for $L_{x}=2\pi$.
The details of the bifurcation diagrams are strongly different but each attractor finally disappears 
in a boundary crisis bifurcation. The Reynolds numbers of the boundary crisis for $1.5\pi$ and $3\pi$ are $781$ and $854$, respectively.

Although, the dynamics of the attractor created in the bifurcation cascade for the case $L_{x}=2\pi$ is chaotic 
the complexity of the flow structure is rather simples due to the symmetry constrain and
the small size of the periodic domain. The merging of attractors
with different spatial wavelength (e.g in a merging crisis) is  a possible scenario that could 
add spatial complexity to the trajectories. In a computational domain
with a streamwise length of  $6\pi$ and a width of $2\pi$ all three attractors or saddles, respectively, 
are present and mergings could actually exist.

\section*{Conclusion}
Subcritical turbulence in plane Poiseuille flow 
is generated in a bifurcation cascade starting at a the simple travelling wave $TW_{Eyz}$.
Although, the travelling wave and the cascade were identified and studied using
a symmetry restricted system, the are also present in the full system but in
an unstable subspace. $TW_{Eyz}$ does not connect to the TS-wave or the laminar flow.
Thus, the state and the boundary formed by the stable manifold of its lower branch, which separates two parts of the state space, can also exist above $Re_{crit}$. 
This might explain why bypass-transition and TS-transition coexist in the supercritical range.
For a detailed understanding of the coexistence of both transitions
the symmetry restriction must be dropped in further studies, because the TS-wave do not exist in the
studied mirror-symmetric subspace. Since coexisting transition scenarios are a phenomenon that also exists in other flows with a base flow that has a linear instability,
such as the Blasius- or the asymptotic suction boundary layer, the results found for PPF can also improve our understanding of these flows.

At low Reynolds numbers PPF \citep{Tuckerman2014} and other subcritical flows \citep[e.g.][]{Prigent2003,Moxey2010,Duguet2010b} show a complex spatio-temporal
behavior . The described existence of saddles with different spatial wavelengths and the possibility of
merging might provide a possible to understand this phenomenon from a dynamical systems point of view.

This work was supported in part by the DFG within FOR 1182.

\bibliographystyle{tsfp9}

\end{document}